\documentclass[backref, breaklinks, twocolumn, colorlinks]{aastex63}   
\hypersetup{linkcolor=blue,citecolor=blue,filecolor=cyan,urlcolor=magenta}

\usepackage{enumitem}
\usepackage{mathtools}
\usepackage{graphicx}
\usepackage{amssymb}
\usepackage{amsmath}
\usepackage{ulem}
\usepackage{epstopdf}
\usepackage{color}
\usepackage{graphicx}
\usepackage{xcolor}
\usepackage{needspace}

{\catcode`\@=11
\gdef\SchlangeUnter#1#2{\lower2pt\vbox{\baselineskip 0pt\lineskip0pt
\ialign{$\m@th#1\hfil##\hfil$\crcr#2\crcr\sim\crcr}}}}
\def\gtrsim{\mathrel{\mathpalette\SchlangeUnter>}}
\def\lesssim{\mathrel{\mathpalette\SchlangeUnter<}}

\begin{document}
\tighten

\title{Fast Radio Burst Trains from Magnetar Oscillations}
\shorttitle{FRB Trains}

\author[0000-0002-9249-0515]{Zorawar Wadiasingh}
\affil{Astrophysics Science Division, NASA Goddard Space Flight Center, Greenbelt, Maryland 20771, USA}
\affil{Universities Space Research Association (USRA) Columbia, Maryland 21046, USA}
\affil{Centre for Space Research, North-West University, Potchefstroom, South Africa }

\author[0000-0003-2759-1368]{Cecilia Chirenti}
\affil{Department of Astronomy, University of Maryland, College Park,
Maryland 20742, USA}
\affil{Astroparticle Physics Laboratory NASA/GSFC, Greenbelt, Maryland 20771, USA}
\affil{Center for Research and Exploration in Space Science and Technology, NASA/GSFC, Greenbelt, Maryland 20771, USA}
\affil{Center for Mathematics, Computation and Cognition, UFABC, Santo André-SP, 09210-170,  Brazil}

\begin{abstract}
Quasi-periodic oscillations inferred during rare magnetar giant flare tails were initially interpreted as torsional oscillations of the neutron star (NS) crust, and have been more recently described as global core+crust perturbations. Similar frequencies are also present in high signal-to-noise magnetar short bursts. In magnetars, disturbances of the field are strongly coupled to the NS crust regardless of the triggering mechanism of short bursts. For low-altitude magnetospheric magnetar models of fast radio bursts (FRBs) associated with magnetar short bursts, such as the low-twist model, crustal oscillations may be associated with additional radio bursts in the encompassing short burst event (as recently suggested for SGR~1935+2154). Given the large extragalactic volume probed by wide-field radio transient facilities, this offers the prospect of studying NS crusts leveraging samples far more numerous than galactic high-energy magnetar bursts by studying statistics of sub-burst structure or clustered trains of FRBs. We explore the prospects for distinguishing NS equation of state models with increasingly larger future sets of FRB observations. Lower $l$-number eigenmodes (corresponding to FRB time intervals of $\sim5-50$ ms) are likely less susceptible than high-$l$ modes to confusion by systematic effects associated with the NS crust physics, magnetic field, and damping. They may be more promising in their utility, and also may corroborate models where FRBs arise from mature magnetars. Future observational characterization of such signals can also determine whether they can be employed as cosmological ``standard oscillators" to constrain redshift, or can be used to constrain the mass of FRB-producing magnetars when reliable redshifts are available.
\end{abstract} 

\section{Introduction}
\label{intro}

Fast radio bursts (FRBs) are radio transients characterized by millisecond durations, brightness temperatures $\gtrsim 10^{30}$ K, extraordinary energetics and high fractional linear polarization. Extragalactic FRBs can be useful probes of the intergalactic medium \citep{2020Natur.581..391M} and other cosmological parameters \citep[e.g.,][]{2018NatCo...9.3833L}. 

In most astrophysical models, the plasma (and associated wave modes) which are involved with the FRB production must be of low entropy\footnote{{The observed ``coherent" radio emission is nonthermal and highly linearly polarized, which demands involvement of nonthermal plasmas and ordered plasma/fields.}}. The inner magnetospheres of neutron stars (NSs), particularly magnetars, are a natural candidate {{\citep[e.g.,][]{2017ApJ...838L..13L,2017MNRAS.468.2726K,2018ApJ...852..140W,2019ApJ...879....4W,2020arXiv200505093L}}}. Indeed, FRB-like events reported by {{the Canadian Hydrogen Intensity Mapping Experiment \citep[CHIME,][]{2020arXiv200510324T} and STARE2 \citep{2020PASP..132c4202B,2020arXiv200510828B}}} associated with a {{hard X-ray}} short burst\footnote{Recurrent {{hard X-ray}} short bursts, of energy $\sim10^{36}-10^{42}$~erg and duration $T_{90}\sim5-500$~ms, are the most numerous {{observed}} type of {{high-energy}} magnetar burst. They are distinguished from giant flares by much lower spectral peaks (typically below 100 keV) and total energetics, and lack of strong pulsating tails/afterglows. {{See \cite{2008A&ARv..15..225M,2015RPPh...78k6901T} for reviews and \cite{2015ApJS..218...11C} for a recent catalog.}}} from SGR~1935+2154 \citep[e.g.,][and references therein]{2020arXiv200506335M,2020arXiv200511071L} suggest that some fraction of extragalactic FRBs originate from mature {{(age $\gtrsim 1$ kyr)}} magnetars {{\citep[for a survey of models, see][]{2020arXiv200505283M}}}. 

The low-twist model is one such magnetospheric magnetar model for FRBs with an explicit connection to hard X-ray short bursts \citep{2019ApJ...879....4W,Wadiasingh2020}. It also proposes that trains\footnote{Or ``sub-bursts", hereafter adopted interchangeably. {{See \S\ref{modeident}, and references therein, for examples.}}} of radio bursts could be associated with strong crustal oscillations. The trigger\footnote{{{That is, the fast instability mechanism that results in individual short bursts, or temporally-correlated clusters of spikes.}}} for {{hard X-ray}} short bursts, and FRBs, may be internal \citep[e.g.,][]{2011ApJ...727L..51P,2017ApJ...841...54T,2019MNRAS.488.5887S} or external {{\citep[e.g.,][]{2012MNRAS.427.1574L,2020arXiv200505093L}}}. In the low-twist model, all FRBs ought to be associated with {{hard X-ray}} short bursts but not conversely owing to low-charge-density conditions necessary for strongly-fluctuating $e^\pm$ pair cascades needed for the pulsar-like emission~\citep{PhysRevLett.124.245101}. In this framework, more prolific repeaters {{\citep[e.g.~FRB~180916,][]{2020arXiv200110275T}}} may be rare mature magnetars with long spin periods \citep[see][for details]{2020MNRAS.tmp.1934B} {{rather than very young hyperactive ones}}. The charge-starvation condition for magnetic $e^\pm$ cascades sets a minimum energy scale which distinguishes FRBs from radio emission from corotationally-driven electric fields in canonical pulsars. Indeed, the FRB-associated short burst in SGR~1935+2154 was spectrally distinct from other bursts in that magnetar\footnote{But more in line with some short bursts in other magnetars \citep[e.g.,][]{2012ApJ...756...54L}.} which did not produce FRBs yet it was unremarkable in light curve structure, temporal evolution or apparent energetics \citep[][]{2020arXiv200611358Y}. This suggests a similar trigger/driver yet with distinct magnetospheric conditions.

Regardless of the trigger's internal/external nature, the magnetic field couples to mobile electrons and more fixed ions in the crust. Disturbances can then excite short-lived characteristic oscillation modes of the NS.

Such quasi-periodic oscillations (QPOs) have been reported in galactic magnetars in both {{hard X-ray}} short bursts \citep{2014ApJ...795..114H,2014ApJ...787..128H} (not unlike those in SGR~1935+2154) and in giant flare tails {{\citep[e.g.,][]{2005ApJ...628L..53I,2005ApJ...632L.111S,2006ApJ...637L.117W,2006ApJ...653..593S,2007AdSpR..40.1446W,2019ApJ...871...95M}}}. Indeed, the two CHIME pulses associated with SGR~1935+2154 are approximately aligned {{(within $\sim 6$ ms), systematically lagging}} with reported hard X-ray peaks \citep{2020arXiv200506335M}, and a third X-ray peak exists apparently at a similar temporal cadence. {{Besides alignment, the peak separation of radio and X-rays is comparable at $\sim 30$ ms, much larger than component widths of $\sim 3$ ms or uncertainties associated with their position.}} QPO-like structure at $\sim35$\,Hz is also suggested in HXMT light curves \citep{2020arXiv200511071L}. Moreover, the radio pulses precede the hard X-ray peaks {{by up to $\sim 6$ ms}} as reported in \cite{2020arXiv200506335M}, disfavoring magnetar models which propose radio emission originating outside the light cylinder or those that trigger the radio after X-rays. 

Sub-bursts have been observed in many FRBs. {{Formally, these are temporal clusters of events, spikes, or multipeak substructure within bursts with much shorter interarrival times than between clusters. Such bimodality in distribution of waiting times is an assumption which appears to be true in both FRBs and hard X-ray short bursts.}} We adopt the conjecture of \cite{2019ApJ...879....4W} that these FRB trains are due to magnetar oscillations. There exists a {{significant gap}} in the waiting time distribution for FRB~121102 between the bulk of recurrences (which exhibit similar {{lognormal}} population properties as magnetar {{hard X-ray}} short bursts) and a minority of short-waiting-time trains \citep[see Fig.~2 in][]{2019ApJ...879....4W}. {{The interarrival time of clusters (i.e. trains) of radio spikes are $\gtrsim 3\sigma$ away from the lognormal mean/peak. The gap suggests trains (i.e. spikes within temporal clusters) are temporally-correlated and share a trigger. Likewise, \cite{2020arXiv200803461C} report that a waiting time of $\sim 40$ ms between spikes in their Effelsberg data is only $3\times 10^{-5}$ probable with Poissonian expectations. Indeed, \cite{2015ApJ...810...66H} also found bimodality (i.e. a gap) in the waiting time distribution of spikes in a hard X-ray short burst storm of SGR 1550--5418.}}

Given the extensive extragalactic volume probed by radio survey facilities, in contrast to the limited detection volume for magnetar short bursts by current GRB instruments \citep[e.g.,][]{2019ApJ...879...40C}, our conjecture offers the prospect of studying NS crusts from samples far larger than afforded by galactic magnetars. Furthermore, the {{spacing and alignment (with a shift of $\sim 6$ ms)}} of X-ray/radio peaks in SGR~1935+2154 suggests that FRBs might be a cleaner probe of the oscillation period than X-rays, owing to their temporal narrowness and high signal-to-noise ratio. {{The crucial point is that the radio and X-rays have a peak-to-peak timescale which are indistinguishable from each other.}}

The current sample of reported FRBs appears insufficient to strongly support or falsify our conjecture. Yet, CHIME and other wide-field transient facilities are expected to imminently report $\gtrsim10^3$ FRBs. Particularly if the magnetar progenitors are similar in mass, more FRB trains might provide strong support for this model. Moreover, if such additional data show that the eigenmodes are standardizable\footnote{That is, if correlations exist between observables which collapse model degeneracies in mode identification.}, this establishes yet another route to estimating redshift of FRBs independent of dispersion measure.

In \S\ref{primer}, we briefly review the relevant physics. In \S\ref{modeident} we present an illustrative case: supposing that burst intervals in reported FRB trains are oscillations, we identify them with specific eigenmodes, adopting two representative NS equations of state (EOS). In \S\ref{outlook} we consider how future observations might be exploited.

\needspace{5\baselineskip}
\section{ {{Magnetar Oscillations - A Primer for Nonspecialists}}}
\label{primer}

\cite{1998ApJ...498L..45D} originally suggested that SGRs could frequently be subjected to starquakes, which would likely excite oscillation modes. Therefore the QPOs observed in the giant flares of SGR 1806--20 and SGR 1900+14 were initially interpreted as torsional crustal modes  {{\citep[e.g.,][]{2005ApJ...628L..53I,2005ApJ...632L.111S,2006ApJ...637L.117W,2006ApJ...653..593S,2007AdSpR..40.1446W,2007MNRAS.374..256S,2008ApJ...680.1398T}}}. Similar identifications of QPOs in SGR~J1550--5418's {{hard X-ray}} short bursts were proposed by \cite{2016NewA...43...80S}. 

The inclusion of a strong magnetic field in the calculation of the oscillations causes small changes in the frequencies of these modes. It also introduces coupling with the continuum of MHD modes in the core and faster damping \citep{2006MNRAS.368L..35L}. 

Longer lived global (core+crust) modes need eigenfrequencies in gaps of the MHD continuum spectrum \citep{2012MNRAS.421.2054G}, which can also be ``broken" by the coupling between axial and polar modes \citep{2012MNRAS.423..811C}, or by tangled magnetic field configurations \citep{2016ApJ...823L...1L}. More sophisticated models have included ingredients such as superfluidity in the study of global oscillations, which also depend in a major way on the details of the crust \citep[see][]{2015RPPh...78k6901T}.
 
 \begin{deluxetable}{c|c|c|c|c|c}
\tablenum{1}
\tabletypesize{\footnotesize}
\tablewidth{\textwidth}
\tablecaption{Adopted EOS Models \label{tab:models}}
\tablewidth{0pt}
\tablehead{
\colhead{Model} &\colhead{Core EOS} & \colhead{Crust EOS} & \colhead{$M$\,[$M_{\odot}$]} & \colhead{$R$\,[km]} & \colhead{$\rho_c$\,[g/cm$^3$]} 
}
\startdata
I & APR\,(1) & Gs\,(3) & 1.4 & 12.4 & $2.01 \times 10^{14}$ \\
 II  & SLy\,(2) & SLy\,(3) \cite{} & 1.4 & 11.7 & $1.34 \times 10^{14}$\\
\enddata
\tablecomments{$\rho_c$ is the crust--core transition density.}
\tablerefs{(1) \cite{1998PhRvC..58.1804A} 
(2) \cite{DouchinHaensel2001} (3) \cite{2012PhRvC..85e5804S} }
\end{deluxetable}

We adopt the simplest model of torsional oscillations of the nonmagnetized NS crust. A more detailed description of global modes can also be straightforwardly applied if desired, but would introduce more assumptions on the NS+field configuration. 
For fundamental ($n=0$) torsional crustal modes with multipole number $l$, the eigenfrequency is approximately proportional to $l$ \citep{2007MNRAS.374..256S}
\begin{equation}
\nu_{l,n=0} \simeq
\frac{\nu_{2,0}}{2}\sqrt{(l-1)(l+2)}\,.
\label{eq:1}
\end{equation}
 The influence of the crustal magnetic field $B$ in the frequencies can be described \citep{1998ApJ...498L..45D,2001MNRAS.328.1161M} by a multiplicative correction,
 \begin{equation}
 \nu^{\rm mag}_{l,n} \simeq \nu_{l,n}\sqrt{1+ \alpha_{l,n}\left(\frac{B}{B_{\mu}}\right)^2}, 
 \label{numag}
 \end{equation}
 where $\alpha_{l,n}$ is a coefficient of order unity \citep{2007MNRAS.375..261S} and $B_{\mu} \approx 4 \times 10^{15}$ G. Spatial inhomogeneity of $B$ within the crust, or time-evolution or rearrangement of $B$ between bursts, may lead to systematic variations of eigenfrequencies over time -- see, for instance observed variation of frequencies below 40~Hz in \cite{2019ApJ...871...95M}. 

The eigenfrequencies depend more strongly on the mass and EOS \citep[especially the crust EOS, see also][]{2014PhRvC..90b5802D}, and more weakly on the $B$ configuration and other details of the NS model. The small number of detections so far and known degeneracies in the modeling make it challenging to solve the inverse problem. However, constraints obtained on the crust EOS would be complementary to  constraints on the (core) EOS from observations of binary NS mergers with LIGO \citep{2018PhRvL.121p1101A} and of PSR~J0030+0451 with NICER \citep{2019ApJ...887L..24M,2019ApJ...887L..21R}. 

\begin{figure*}[htp]
\centering
\includegraphics[width=0.95\textwidth]{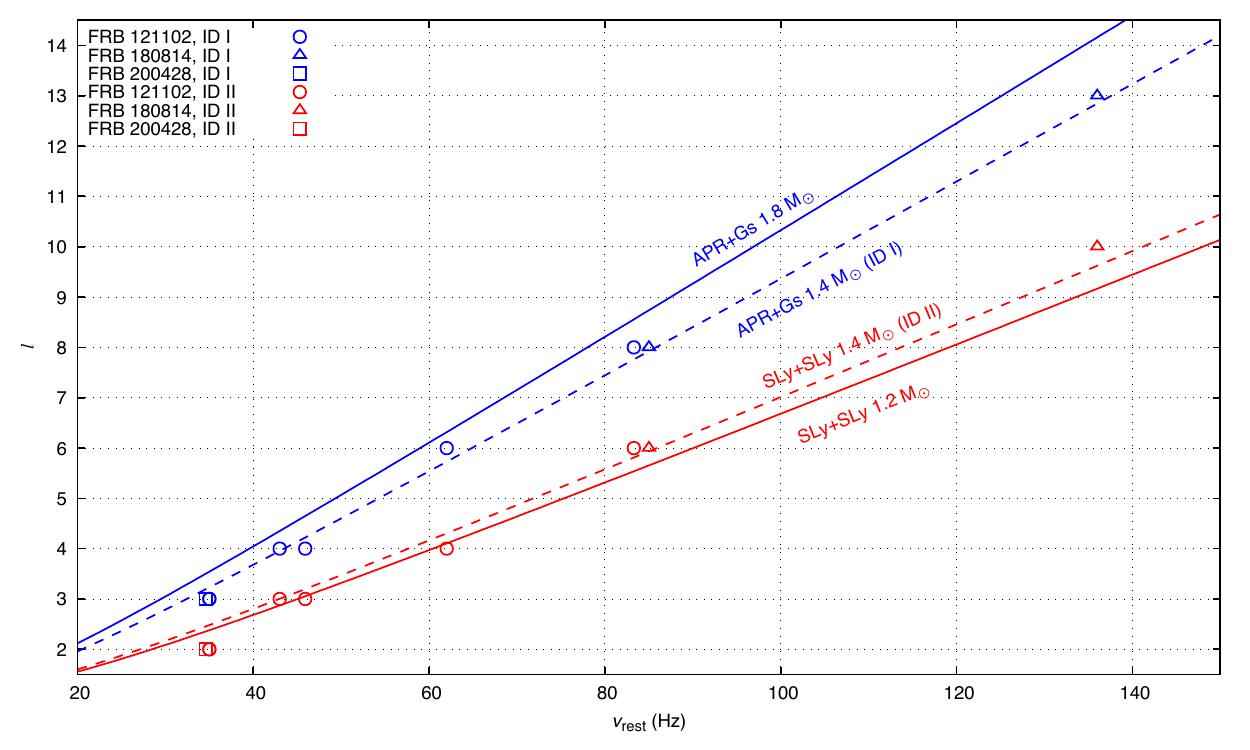}
\caption{Mode identifications with EOS models I and II (Table~\ref{tab:models}). Low ($1.2M_\odot$\,\,Sly+Sly) and high mass ($1.8M_\odot$\,\,APR+Gs) cases bracket possible systematic variation due to EOS and unknown mass.}
\label{fig1}
\end{figure*}

Investigations of the parameter space demonstrate that torsional eigenfrequencies for each mode $l$ decrease with increasing total mass of the NS (with a relative variation of $\sim30\%$ from $1\rightarrow2 M_{\odot}$). However, they increase for harder crust EOSs ($\sim30\%$ relative variation at $1.4M_{\odot}$ across different models). There is an additional (but weaker) effect of the core EOS: the frequencies decrease for harder core EOSs, with $\sim5\%$ relative variation at $1.4M_{\odot}$, for softer EOSs consistent with current LIGO constraints. For a range of masses ($1-2M_{\odot}$) and EOSs (both crust and core) the eigenfrequencies at fixed $l$ can vary $\sim60\%$ (see Figure~\ref{fig1}). For example, \cite{2019PhRvD.100d3017D} found that $\nu_{2,0} \sim 18-30$\,Hz, with increasing values for harder crusts. 

Note that model eigenfrequencies account for the NS's gravitational redshift, and the quoted values are for a distant observer at rest. For the FRB context, a factor associated with cosmological redshift is necessary (see \S\ref{modeident}).

The damping times are much more model dependent and vary more strongly with the details of the crust $B$ configuration. Coupling with MHD modes in the core can shorten the damping time, which is estimated to be roughly from $\mathcal{O}(1 \,\rm{ms})-\mathcal{O}(1 \,\rm{s})\propto$~$B^{-1/2}$ \citep[e.g.,][]{2006MNRAS.368L..35L,2012MNRAS.421.2054G}. This is consistent with the reanalysis of SGR 1806--20 data reported by \cite{2019ApJ...871...95M} {and with the analysis of its 625 Hz QPO performed by \cite{2014ApJ...793..129H}}. 
The duration of FRB trains may constrain the $B$ strength, although it is very configuration dependent. For instance, \cite{2012MNRAS.421.2054G} find that for a dipole with $B \gtrsim$ few$\times 10^{14}$~G, the damping time would be so short such multiple observed oscillation periods are unlikely. Therefore, observation of trains of low $l$-number modes may suggest a mature magnetar disfavoring models which invoke extreme young magnetars (age $\ll 1$ kyr and $B\sim 10^{16}$ G) as FRB progenitors. Fortuitously, $B$ corrections to nonmagnetic eigenfrequencies would also be much smaller in mature magnetars.

\begin{deluxetable*}{c|CC|CC|c}
\tablenum{2}
\tablecaption{Reported Trains in FRB~121102 \label{tab:nus121102}}
\tablewidth{0pt}
\tablehead{
\colhead{Burst} & \colhead{$\nu_{\rm obs}$ [Hz]} & \colhead{$\nu_{\rm rest}$ [Hz]} & \colhead{Mode Identification I} & \colhead{Mode Identification II} & \colhead{Refs.} 
}
\startdata
8$\rightarrow$9 & 0.0143^{-1} & 83.3 & n=0, l=8 & n=0, l=6 & (1)\\
27$\rightarrow$28 & 0.00522^{-1} & 228.3 & n=0, l=22 & n=0, l=16 & (1)\\
28$\rightarrow$29 & 0.00195^{-1} & 612.2 & n=1 & n=1 & (1)\\
30$\rightarrow$31 & 0.01925^{-1} & 62.0 & n=0, l=6 & n=0, l=4 & (1)\\
68$\rightarrow$69 & 0.00242^{-1} & 492.0 & n=0, l=47 & n=0, l=36 & (1)\\
81$\rightarrow$82 & 0.00267^{-1} & 446.4 & n=0, l=43 & n=0, l=32 & (1)\\
\hline
B5$\rightarrow$B6 & 0.108^{-1} & 11.0 &- & - & (2)\\
B35$\rightarrow$B36 & 0.026^{-1} & 45.9 & n=0, l=4 & n=0, l=3 & (2)\\
\hline
``Figure 4" & 0.034^{-1} & 35 & n=0, l=3 & n=0, l=2 & (3)\\
\hline
03 & 0.028^{-1} & 43 & n=0, l=4 & n=0, l=3 & (4)\\
05 & 0.034^{-1} & 35 & n=0, l=3 & n=0, l=2 & (4)\\
\enddata
\tablecomments{We adopt $z=0.19273$ \citep{2017ApJ...834L...7T}.} 
\tablerefs{(1) \cite{2018ApJ...866..149Z} 
(2) \cite{2019ApJ...877L..19G}
(3) \cite{2017MNRAS.472.2800H}
(4) \cite{2020arXiv200608662C}.}
\end{deluxetable*}

\begin{deluxetable*}{c|CC|CC}
\tablenum{3}
\tablecaption{Sub-bursts Reported in FRB~180814.J0422+73 \label{tab:nus180814}}
\tablewidth{0pt}
\tablehead{
\colhead{Burst} & \colhead{$\nu_{\rm obs}$ [Hz]} & \colhead{$\nu_{\rm rest}$ [Hz]} & \colhead{Mode Identification I} & \colhead{Mode Identification II} 
}
\startdata
``09/17"  & 0.013^{-1} & 85 & n=0, l=8 & n=0, l=6 \\
``10/28" & 0.0081^{-1} & 136 & n=0, l=13 & n=0, l=10 \\
\enddata
\tablecomments{{``09/17" and ``10/28" refer to two bursts with multiple resolved sub-bursts -- see Extended Data Table~1 and Figure~1 in \cite{2019Natur.566..235C}.}}
\end{deluxetable*}

\needspace{2\baselineskip}
\section{Tentative Mode Identification in FRBs}
\label{modeident}

As an illustrative first step, we consider time intervals\footnote{{This is quite different from what has been done in the QPO analysis of X-ray light curves. Our conjecture that these time intervals are related to oscillations is based on the temporal correlation between X-ray light curve features and radio bursts in SGR 1935+2154 described in \S \ref{intro}.}} between reported trains or sub-bursts. The distinction between sub-bursts within longer FRBs and well-separated FRB trains is considered physically not meaningful, since instrumental threshold and scatter-broadening of trains can influence such categorization. {{For instance, two well-separated spikes within a short time interval may be categorized as separate bursts (train) or a single burst with two sub-bursts, and can be influenced by how DM is optimized \citep{2019ApJ...876L..23H}, instrumental threshold which sets the baseline for any ``jagged iceberg" signal, and software pipelines.}} Some fine structure within bursts could also result from pair cascade nonstationarity \citep{2010MNRAS.408.2092T,2013MNRAS.429...20T}, lensing by compact objects \citep[e.g.,][]{2020arXiv200212533S,2020PhRvD.102b3016L}, or high crustal $n \geq 1$ modes. Alternatively, fine sub-burst structure could also result from propagation effects by strongly-inhomogeneous scattering and scintillation \citep[e.g.,][and references therein]{2019ARA&A..57..417C}.

{{To be clear: we adopt time interval between FRB spikes as a proxy for a putative magnetar oscillation period. This provides an illustrative example, consistent with the correlation between X-rays and radio spikes in SGR 1935+2154. A large sample of FRB radio pulses (see \S\ref{outlook}) and/or more detailed analysis of corresponding X-ray data (similar to those which found QPOs in other magnetar short bursts) is necessary to confirm our hypothesis but is beyond the scope of this paper.}}

Longer timescale variability ($\gg 1$ ms) which cannot easily be ascribed to propagation effects (without invoking contrived plasma screens or emission regions far away from the NS) are likely more secure for potential identification with crustal eigenmodes. Thus we focus on these for tentative mode identification. However, we emphasize frequencies obtained from the intervals between bursts are a crude estimate that must be confirmed with a more rigorous analysis similar to that of \cite{2019ApJ...871...95M}.

Reporting of FRB trains is not uniform in the current literature. In particular, \cite{2019ApJ...885L..24C,2020ApJ...891L...6F,2020arXiv200110275T} report several sub-bursts in various FRBs mostly commensurate with those that we consider in this preliminary work, but accurate time intervals between those components are not detailed. 

We adopt two NS models for candidate eigenmode identification in FRB trains--see Table~\ref{tab:models} and Appendix~\ref{appendix_ident}. Our choice is guided by current constraints on the radius $R$ of $1.4 M_{\odot}$ NSs from GW170817 by \cite{2018PhRvL.121p1101A} ($R\lesssim 13.5$\,km) and by {\it{NICER}} inferences for PSR~J0030+0451 \citep[][$R \approx13\pm1$\,km]{2019ApJ...887L..24M,2019ApJ...887L..21R}.

Given the cosmological nature of FRBs, candidate frequencies $\nu_{\rm obs}$ must be transformed to the comoving inertial rest frame of the host galaxy at redshift $z$ by
\begin{equation}
\nu_{\rm rest} =  \nu_{\rm obs}(1+z)
\label{redshift}
\end{equation}
for comparison with model eigenfrequencies.

\needspace{2\baselineskip}
\subsection{FRB~200428 and SGR~1935+2154}

FRB-like bursts temporally coincident with hard X-rays from SGR~1935+2154 support our conjecture that FRB trains may carry an imprint of the progenitor crustal dynamics. \cite{2020arXiv200506335M} in fact report that there are three X-ray peaks roughly separated at $\sim 30$ ms, leading to the intriguing possibility that these peaks result from crustal oscillations. {{Indeed, such correlation of radio bursts and features of the X-ray light curve suggests these features do not arise from a red noise process. Future radio/X-ray bursts may clarify this view.}} This motivates comparison of time intervals with crustal oscillation periods in other FRBs.  
The $28.9$\,ms time interval (much larger than the reported scattering time $\sim0.8$\,ms) between the CHIME bursts \citep{2020arXiv200510324T} corresponds to $\nu_{\rm obs} \approx 34.6$\,Hz. The eigenmode identification thus is $n=0,l=3$ or $n=0,l=2$ (see Figure \ref{fig1}) at $z\simeq0$.
An alternative scintillation scenario has been proposed for SGR~1935+2154 \citep{2020arXiv200613184S}, but this model is incompatible with a magnetospheric emission scenario.

\needspace{2\baselineskip}
\subsection{FRB~121102}

FRB~121102 is one of the most well-studied recurrent FRBs, and the first to be localized with a redshift. Hundreds of bursts have been reported since its discovery, including a ``storm" in 2017 which emitted 93 bursts \citep{2018ApJ...866..149Z} over $\sim5$~hours. For the vast majority of the bursts in that storm, the interarrival times are lognormally distributed with a mean of $\sim60$ s and width $0.7$\,dex. A separate, smaller, population of the 93 bursts have short interarrival times, listed in Table~\ref{tab:nus121102}.

FRB~121102 also exhibits complex time-frequency structures in time-resolved analysis \citep[e.g.,][]{2019ApJ...876L..23H}. These structures correspond to variability at frequencies $\gtrsim600$\,Hz {{(i.e., those associated with largest timescales in their Fig. 3 top left panel)}}. Local galactic diffractive interstellar scintillation can account for some fine-structure, but not for longer timescales {{considered in this work}}. 

\cite{2018ApJ...866..149Z} searched for periodicities in the arrival times of bursts in their $\sim5$ hour window and did not find any compelling signals {{for long-lived periodicity}}. Yet, candidate periods quoted by \cite{2018ApJ...866..149Z} are compatible with some of the candidate frequencies reported in Table~\ref{tab:nus121102}. If the oscillations are quickly damped in the signal (and possibly re-excited) the search for QPOs must focus on shorter segments of data \citep{2019ApJ...871...95M}.

We also consider other burst intervals reported in the literature in Table~\ref{tab:nus121102} for $\gg1$ ms timescale trains. For some burst intervals in Table~\ref{tab:nus121102}, unseen intervening bursts could exist, e.g., if the magnetosphere is sufficiently polluted and strong nonstationary $e^\pm$ cascades are quenched. Thus the table comprises minimum frequencies (with the real crustal eigenfrequency an integer multiple $i=1,2,3,...$ larger). This may explain the largest time interval B5\,$\to$\,B6 in Table~\ref{tab:nus121102}, which prevents mode identification.

We see that most of these candidate modes are compatible with those inferred in galactic magnetars\footnote{The details of the initial perturbation(s) likely select which modes are excited with detectable amplitude \citep{2017APS..APR.Y4007B}.}.
The lower $l$ modes corresponding to $\nu_{\rm rest}\sim35-45$~Hz suggests that damping times are not short, i.e., FRB~121102 is compatible with mature magnetar with a relatively moderate $10^{13}$~G $\lesssim B\ll10^{16}$~G.
The higher frequencies quoted in Table~\ref{tab:nus121102} can be tentatively identified with larger $l$-numbers. The interpretation of these modes is unclear, and might relate to oscillations that involve only a small area of the crust. A larger sample is necessary to establish discreteness in the spectrum of modes. Importantly, a more rigorous analysis is necessary to identify possible high-frequency QPOs in the data. Given our preliminary analysis, it is therefore likely that not all frequencies quoted here will be replicated. 

\needspace{2\baselineskip}
\subsection{FRB 180814.J0422+73}

FRB~180814.J0422+73 is a prominent repeater \citep[][]{2019Natur.566..235C}. \cite{2019Natur.566..230C} measure a characteristic scattering time $<0.4$ ms. 
The ``9/17" sub-burst in FRB~180814.J0422+73 \citep{2019Natur.566..235C} is so strikingly regular that it has been proposed to be associated with a spin period \citep{2020ApJ...890..162M}. Yet, it is also broadly consistent with crustal modes observed in magnetars. Our mode identification in Table~\ref{tab:nus180814} adopts $z=0.1$, based on arguments in \cite{2019Natur.566..235C}. Alternatively, adopting models I and II, the redshift is estimated as $z\sim0.11-0.14$.

The $\sim50$\,ms duration of the ``9/17" train also suggests the oscillation damping time is not short and FRB~180814.J0422+73 arises from a mature magnetar.

\needspace{3\baselineskip}
\section{Outlook for Standardizing FRB Trains}
\label{outlook}

\begin{figure*}[t]
\centering
\includegraphics[width=0.999\textwidth]{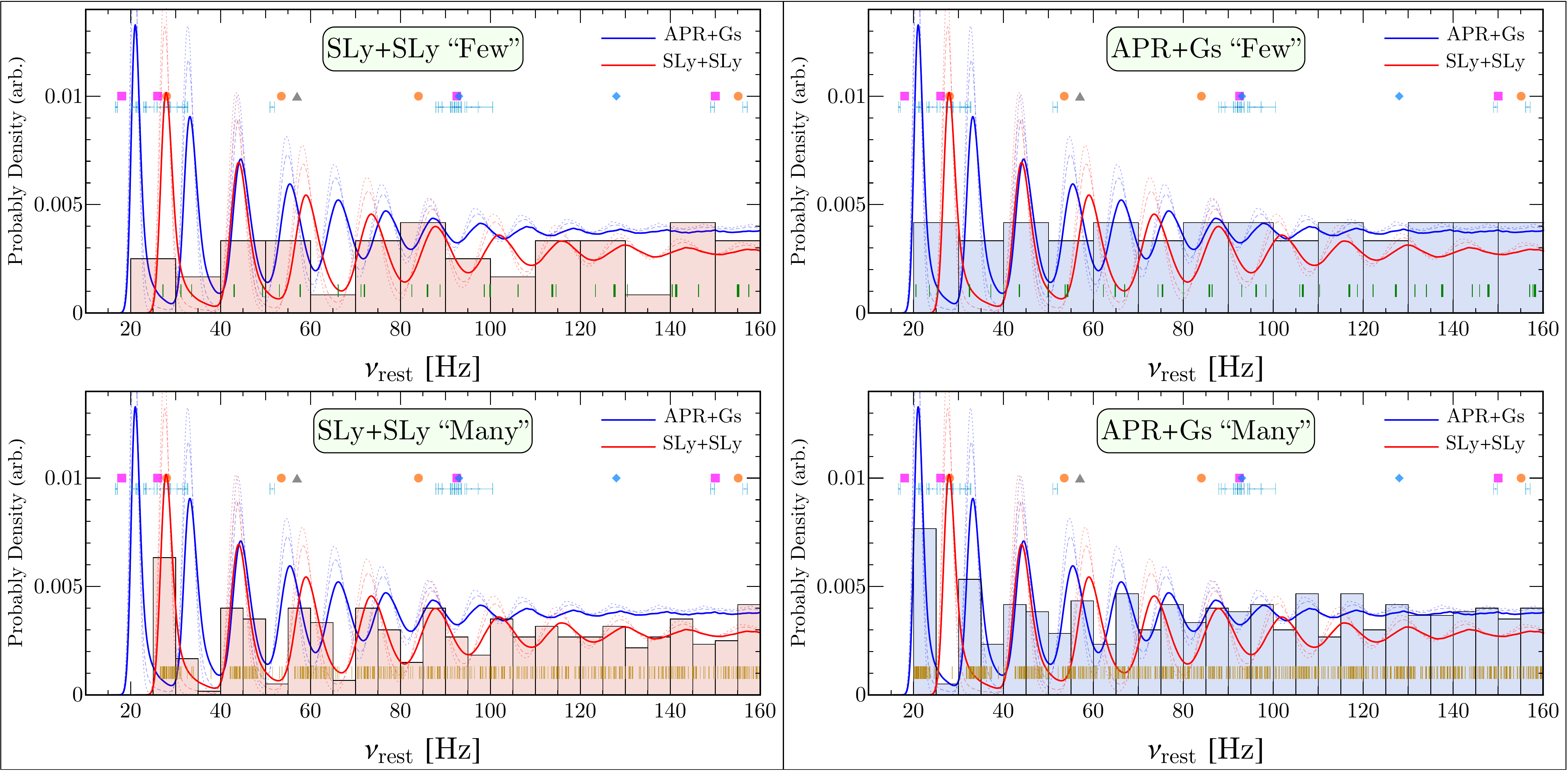}
\caption{ Realizations from the model PDF (red and blue curves) of putative crustal oscillation frequencies from an unbiased sample of magnetars of varying modeled mass and magnetic field if, conservatively, all $l$-numbers were excited/observed with equal probability. Upper (``Few") and lower (``Many") panels depict $N= 5$ and $N=50$ samples per $l$-number in the model with the associated histogram (green and brown vertical lines depict individual members of histogram bins). Of order $\sim 50$ frequency measurements per $l$-number are necessary to distinguish deviations from uniformity below 100 Hz at $2-3\sigma$ confidence. Extant reports of QPOs in SGRS are also shown, which highlight not all $l$-numbers may be uniformly produced -- SGR 1900+14 (\cite{2005ApJ...632L.111S} orange circles), SGR 1806--20 (\cite{2006ApJ...653..593S} magenta squares, \cite{2014ApJ...795..114H} gray triangle, \cite{2019ApJ...871...95M} cyan fences) and SGR 1550--5418 (\cite{2014ApJ...787..128H} blue diamonds). See Appendix~\ref{appendixcavaets} for identification of dashed and dotted lines.}
\label{fig2}
\end{figure*}

Our conjecture is that temporally closely-separated FRBs (i.e. trains or sub-bursts) are associated with crustal oscillations. A crucial point is that such crustal eigenmodes are discrete and follow roughly integer ratios for any individual NS. Additionally, they are dependent only on the characteristics of the NS (such as the total mass, crust EOS and $B$), 
i.e., they are independent of any initial perturbations or transients. 

Based on the reported approximate alignment of radio bursts and hard X-ray peaks in SGR~1935+2154, we propose that the radio can be potentially more advantageous than the X-ray for eigenfrequency identification. The radio can also probe a far larger cosmological volume of bursts. Therefore it is essential that time intervals between sub-bursts be reported by the radio community, barring a more rigorous QPO analysis.

For any individual magnetar, there is likely some additional spread in the candidate eigenfrequencies owing to inhomogeneity and variation of $B$ in the crust. Empirically, this is the case in at least one galactic magnetar \citep[SGR 1806--20,][]{2019ApJ...871...95M}. A population ensemble will also introduce dispersion in candidate train eigenfrequencies due to a variety of factors such as varying progenitor NS masses, crustal $B$ fields, redshifts, beats\footnote{In SGR 1806--20, however, multiple independent eigenmodes are apparently not simultaneously excited in the analysis of \cite{2019ApJ...871...95M}.}, and propagation effects. Yet concentrations, or bands, could be revealed after a redshift correction (for instance, based on dispersion measure) since the influence of NS mass is $<30\%$ if FRBs are produced by mature magnetars with moderate crust $B \lesssim B_\mu$. 

Therefore we can expect a fractional frequency 
spread for each ($n=0$) $l$-mode for the FRB population to be
\begin{equation}
\frac{\Delta \nu_{{\rm rest},l}}{\langle \nu_{{\rm rest},l} \rangle} \approx c_0 \frac{\Delta M}{\langle M \rangle} + c_1 \frac{\Delta B}{ \langle B \rangle} + ...
\end{equation}
where the coefficients $c_i$ are expected to be small (see Figure \ref{fig1}) and will be determined by the EOS and field configuration. Furthermore, such a spread could be asymmetric owing to influences of the field (particularly at higher $l$-numbers).

The relative distribution width (and skewness) in candidate frequencies of the FRB population then determines which $l$-numbers can be differentiated, since eigenfrequencies scale approximately linearly with $l$ while systematics associated with the NS mass, crust $B$ and redshift are multiplicative. Lower $l$-numbers (with $n=0$) are then less prone to such systematic effects and could most easily exhibit integer scaling associated with discreteness. This is readily apparent in Figure~\ref{fig1}. For a fiducial systematic fractional spread of $\sim 20-30\%$, $l$-numbers up to $l\sim6-8$ may be differentiated from neighboring modes if all modes are equally likely in FRB production. However, observations from SGRs indicate that some modes may be skipped (or excited with very low amplitudes) in any given event, 
depending on the details of their initial excitation.

To quantify the above statements, we construct a simplified simulation of a population of candidate magnetar oscillation frequencies (assumed corrected to the rest frame of the source). This may be used to infer discreteness in the distribution and distinguish between oscillation models (in the case highlighted, explicitly EOS models for crustal oscillations). In Figure~\ref{fig2}, we display realizations of such a simulation where we sample $l$-numbers from magnetars of different masses and magnetic fields. The number of realizations, i.e. observed frequencies, then determines the robustness of how well observations can exhibit frequency clustering or distinguish between models.

The model probability density function (PDF) in Figure~\ref{fig2} is shown in the solid red and blue curves for the SLy+SLy and APR+Gs models, respectively. A realization of samples of this PDF with 5 (``Few") and 50 (``Many") frequencies per $l$-number (from the assumed magnetar population) are shown in the upper and lower panels, respectively. See Appendix~\ref{appendixcavaets} for details and caveats for the model.

In the construction of Figure~\ref{fig2}, to highlight possible confusion of identification, we conservatively assume a uniform probability for the spectrum of $l$-modes observed (up to an arbitrarily large $l$-number unimportant for our demonstration) rather than the more specific clustering exhibited in SGRs (plotted markers in Figure~\ref{fig2}). For instance, in SGR 1806--20 \citep[e.g.,][]{2007AdSpR..40.1446W,2019ApJ...871...95M} it appears limited $l$ modes (most often near $\sim20-40$\,Hz and $\sim80-100$\,Hz) are more often present over others, so differentiation in a population may be also possible at higher $l$ modes provided such gaps exist. Interestingly, the $\sim80-85$\,Hz mode is apparently present in both FRB~121102 and FRB 180814.J0422+73, perhaps indicating they are similar in mass. Curiously, there is also apparent concordance of $\sim 20-40$\,\,Hz modes in FRB~121102, SGR~1935+2154 and SGR 1806--20.

As noted above, clustering is readily apparent at lower $l$-numbers in the lower panels with higher number of putative observed frequencies. Frequency spacing in the APR+Gs model is narrower than SLy+SLy, resulting in more confusion at equal $l$-number but also a path for model/EOS discrimination at lower $l$-number via measurement of the spacing of peaks in the histograms in Figure~\ref{fig2}. Comparison of the observed realizations against a uniform distribution with Kolmogorov-Smirnov (KS) and Anderson-Darling (AD) tests can quantify the statistical significance of clustering. Unsurprisingly, we find that the ``Few" cases are statistically indistinguishable from a uniform distribution. In contrast, the ``Many" cases for SLy+SLy and APR+Gs are $\gtrsim3\sigma$ (AD/KS null hypothesis $p \sim 10^{-3}$) and $\gtrsim 2\sigma$ (AD/KS null hypothesis $p \sim 10^{-2}$) away from uniform below 100 Hz, respectively. Significance rises (lowers) if the upper-limit frequency of comparison against a uniform distribution (of the same range of comparison) is lowered (raised) over the adopted 100 Hz.

Figure~\ref{fig2} is a pessimistic illustrative case where $l$-numbers are equally likely without gaps, and there is significant dispersion in the distribution of magnetar masses -- this makes clustering of modes more challenging to distinguish from noise. Observed lower $l$-numbers bursts also must have longer damping times (and therefore lower $B$), thus we likely overestimate eigenfrequency distribution skewness associated with $B \gtrsim B_\mu$. The existence of eigenfrequency gaps can only be constrained observationally, over a unbiased population of FRB trains. From the conservative uniform-$l$ simulation, we conclude that ${\cal O} (10^2)$ unbiased rest-frame corrected frequencies are sufficient to corroborate magnetar oscillations are involved in FRB production.

The standarizability of FRBs depends on the population characteristics of FRBs (or repeater subpopulations) and how well the EOS is known (plausibly, only one EOS describes all NSs). If the observed dispersion across a population of FRBs is small, then trains in FRBs from unknown redshift may be assigned tentative probabilities for different $l$-number, resulting in a probable redshift. Machine learning techniques, over full FRB time-frequency data, may be useful in this goal. 

Alternatively, if redshifts are reasonably well constrained via other methods, then population discrete eigenmode identification can begin constraining the NS EOS. A framework for pooling different astrophysical information for EOS constraints is presented by \cite{2020ApJ...888...12M}. If confirmed, the eigenfrequencies from FRBs can augment a similar analysis. They could provide valuable input on the crustal EOS and usher in a new era for the study of cold dense matter.

\vspace{-0.5cm}
\acknowledgements

\noindent We thank Paz Beniamini and George Younes for useful discussions. We also thank Cole Miller, Kostas D. Kokkotas and Arthur Suvorov for valuable feedback on an earlier version of this manuscript. Z.W. is supported by the NASA Postdoctoral Program. C.C. acknowledges support by NASA under award number 80GSFC17M0002 and by the Brazilian National Council for Scientific and Technological Development (CNPq grant 303750/2017-0). This work has made use of the NASA Astrophysics Data System.

\appendix

\section{An Algorithm for Mode Identification and Parameter Estimation in Individual Sources}
\label{appendix_ident}

Mode identification of observed candidate frequencies in individual sources requires adopting a model (with associated parameters) for eigenmodes at fixed cosmological redshift. In general, for a chosen model and set of parameters, the predicted model $l$-number, $\ell^{\rm model}$, for a candidate observed frequency will not result in an integer value. Let us denote the nearest integer to $\ell^{\rm model}$ as $l^{\rm nearest}$. For instance, inversion of Eq.~(\ref{eq:1}) at $B\ll B_\mu$ at given redshift~$z$ for an observed frequency $\nu_{\rm obs}$ yields candidate $l$-number $\ell^{\rm model} = \left( \sqrt{(3 \nu_{2,0})^2+(4 \nu_{\rm obs} [1+z])^2}/\nu_{2,0} -1 \right)/2$. The identifications in Tables~\ref{tab:nus121102}--\ref{tab:nus180814} are $l^{\rm nearest}$ for the adopted models in Table~\ref{tab:models}. 

This rounding aspect can be used to constrain model parameters or redshift in individual sources (but in practice may not be too constraining owing to poor knowledge of models and source parameters -- a more productive path may be unbiased populations of FRB trains -- see Figure~\ref{fig2}). For example, for a collection of candidate frequencies $i=1 \ldots N$ with weights $w_i$, the quantity $W$ (adopting standard assumptions)
\begin{equation}
W \equiv \sum_i w_i (\ell_{i}^{\rm model} - l_{i}^{\rm nearest})^2 \qquad , \qquad l_{i}^{\rm nearest} = \left \lceil \frac{ \lfloor 2 \, \ell_{i}^{\rm model} \rfloor}{2} \right\rceil
\end{equation}
may be minimized (or sampled) over some set of model parameters to yield constraints (or posterior distributions) for quantities such as mass and redshift. Here $\lfloor x \rfloor$ and $\lceil x \rceil$ are floor and ceiling functions, respectively.

\section{Details and Caveats on the Simulated Models}
\label{appendixcavaets}

For construction of Figure~\ref{fig2}, we assume the masses of magnetars are roughly commensurate with the nonrecycled neutron star population \citep{2012ApJ...757...55O,2013ApJ...778...66K}, and adopt a Gaussian distribution $M = 1.35 \pm 0.15 M_\odot$. For concreteness, we take crustal oscillations models in \cite{2019PhRvD.100d3017D}, interpolated on a grid of masses between $0.8-2.0 M_\odot$ for $\nu_{2,0}$ and calculate $\nu_{l,0}$ via Eq.~(\ref{eq:1}). A higher or lower dispersion in masses can strongly influence the width of peaks in the PDF in Figure~\ref{fig2}. For expediency, we assume Eq~(\ref{numag}) with $\alpha_{l,0} \equiv 1$ and $B_\mu = 4 \times 10^{15}$ G. For the magnetic field distribution in the population of magnetars, we adopt the well-known phenomenological field decay paradigm of \cite{2000ApJ...529L..29C} extended in \cite{2012MNRAS.422.2878D,Beniamini2019} with steady-state distribution $dN/dB \propto B^{-(1+\alpha)}$ for a constant birth rate, with field evolution parameter $\alpha$. For birth magnetic fields, we assume $3\times 10^{14} - 3\times 10^{15}$ G, a identical construction as the ``II" case in \cite{Wadiasingh2020}. The value of $\alpha$ is generally constrained $-1 < \alpha < 1$ by \cite{Beniamini2019} -- we adopt $\alpha = -1$ as the fiducial case (however values of $\alpha = \{0,1\} $ are shown in Figure~\ref{fig2} in dash and dotted lines, respectively). This choice also biases samples to magnetars with higher B over other $\alpha$ values, which highlights possible skewness of frequency clustering. For a physical model associated with $\alpha \sim -1$, see \cite{Beloborodov&Li16}.

There are many caveats associated with such a simple exercise, particularly related to assumed parameters, the model and systematics of sample biases. For instance, the dispersion of masses in magnetars is totally unknown -- no measurement of the mass of a magnetar exists. Nevertheless, with the advent of relatively unbiased scanning wide-field radio survey instruments gathering thousands of FRBs, it is conceivable that various data selection criteria (e.g. on FRB recurrence rates, or exposure time) to minimize possible biases could be implementable.  Rest frame correction over a large population of FRBs could also be feasible with a redshift-DM relation \citep{2020Natur.581..391M}. This is beyond the scope of this work. Model selection, constraints, or falsification, are obviously also possible via standard techniques. For instance, KS and AD tests rule out to $> 2\sigma$ confidence that the ``Many" realization of SLy+SLy is indistinguishable from the APR+Gs realization below 100 Hz. Alternatively, histogram peak-to-peak measurements could generically constrain the slope a model/EOS must follow in Figure~\ref{fig1}.

\bibliographystyle{aasjournal}
\bibliography{magnetarFRBrefs}

\begin{thebibliography}{}
\expandafter\ifx\csname natexlab\endcsname\relax\def\natexlab#1{#1}\fi
\providecommand{\url}[1]{\href{#1}{#1}}
\providecommand{\dodoi}[1]{doi:~\href{http://doi.org/#1}{\nolinkurl{#1}}}
\providecommand{\doeprint}[1]{\href{http://ascl.net/#1}{\nolinkurl{http://ascl.net/#1}}}
\providecommand{\doarXiv}[1]{\href{https://arxiv.org/abs/#1}{\nolinkurl{https://arxiv.org/abs/#1}}}

\bibitem[{{Abbott} {et~al.}(2018)}]{2018PhRvL.121p1101A}
{Abbott}, B.~P., {et~al.} 2018, \prl, 121, 161101,
  \dodoi{10.1103/PhysRevLett.121.161101}

\bibitem[{{Akmal} {et~al.}(1998){Akmal}, {Pandharipande}, \&
  {Ravenhall}}]{1998PhRvC..58.1804A}
{Akmal}, A., {Pandharipande}, V.~R., \& {Ravenhall}, D.~G. 1998, \prc, 58,
  1804, \dodoi{10.1103/PhysRevC.58.1804}

\bibitem[{Amiri {et~al.}(2020)Amiri, Andersen, Bandura, Bhardwaj, Boyle, Brar,
  Chawla, Chen, Cliche, Cubranic, Deng, Denman, Dobbs, Dong, Fandino, Fonseca,
  Gaensler, Giri, Good, Halpern, Hessels, Hill, H{\"o}fer, Josephy, Kania,
  Karuppusamy, Kaspi, Keimpema, Kirsten, Landecker, Lang, Leung, Li, Lin,
  Marcote, Masui, Mckinven, Mena-Parra, Merryfield, Michilli, Milutinovic,
  Mirhosseini, Naidu, Newburgh, Ng, Nimmo, Paragi, Patel, Pen,
  Pinsonneault-Marotte, Pleunis, Rafiei-Ravandi, Rahman, Ransom, Renard,
  Sanghavi, Scholz, Shaw, Shin, Siegel, Singh, Smegal, Smith, Stairs,
  Tendulkar, Tretyakov, Vanderlinde, Wang, Wang, Wulf, Yadav, Zwaniga, \&
  Collaboration*}]{2020arXiv200110275T}
Amiri, M., Andersen, B.~C., Bandura, K.~M., {et~al.} 2020, Nature, 582, 351,
  \dodoi{10.1038/s41586-020-2398-2}

\bibitem[{{Beloborodov} \& {Li}(2016)}]{Beloborodov&Li16}
{Beloborodov}, A.~M., \& {Li}, X. 2016, \apj, 833, 261,
  \dodoi{10.3847/1538-4357/833/2/261}

\bibitem[{{Beniamini} {et~al.}(2019){Beniamini}, {Hotokezaka}, {van der Horst},
  \& {Kouveliotou}}]{Beniamini2019}
{Beniamini}, P., {Hotokezaka}, K., {van der Horst}, A., \& {Kouveliotou}, C.
  2019, \mnras, 487, 1426, \dodoi{10.1093/mnras/stz1391}

\bibitem[{{Beniamini} {et~al.}(2020){Beniamini}, {Wadiasingh}, \&
  {Metzger}}]{2020MNRAS.tmp.1934B}
{Beniamini}, P., {Wadiasingh}, Z., \& {Metzger}, B.~D. 2020, \mnras,
  \dodoi{10.1093/mnras/staa1783}

\bibitem[{{Bochenek} {et~al.}(2020{\natexlab{a}}){Bochenek}, {McKenna},
  {Belov}, {Kocz}, {Kulkarni}, {Lamb}, {Ravi}, \&
  {Woody}}]{2020PASP..132c4202B}
{Bochenek}, C.~D., {McKenna}, D.~L., {Belov}, K.~V., {et~al.}
  2020{\natexlab{a}}, \pasp, 132, 034202, \dodoi{10.1088/1538-3873/ab63b3}

\bibitem[{{Bochenek} {et~al.}(2020{\natexlab{b}}){Bochenek}, {Ravi}, {Belov},
  {Hallinan}, {Kocz}, {Kulkarni}, \& {McKenna}}]{2020arXiv200510828B}
{Bochenek}, C.~D., {Ravi}, V., {Belov}, K.~V., {et~al.} 2020{\natexlab{b}},
  arXiv e-prints, arXiv:2005.10828.
\newblock \doarXiv{2005.10828}

\bibitem[{{Bretz} {et~al.}(2017){Bretz}, {van Eysden}, \&
  {Link}}]{2017APS..APR.Y4007B}
{Bretz}, J., {van Eysden}, A., \& {Link}, B. 2017, in APS Meeting Abstracts,
  Vol. 2017, APS April Meeting Abstracts, Y4.007

\bibitem[{{Caleb} {et~al.}(2020){Caleb}, {Stappers}, {Abbott}, {Barr},
  {Bezuidenhout}, {Buchner}, {Burgay}, {Chen}, {Cognard}, {Driessen}, {Fender},
  {Hilmarsson}, {Hoang}, {Horn}, {Jankowski}, {Kramer}, {Lorimer}, {Malenta},
  {Morello}, {Pilia}, {Platts}, {Possenti}, {Rajwade}, {Ridolfi}, {Rhodes},
  {Sanidas}, {Serylak}, {Spitler}, {Townsend}, {Weltman}, {Woudt}, \&
  {Wu}}]{2020arXiv200608662C}
{Caleb}, M., {Stappers}, B.~W., {Abbott}, T.~D., {et~al.} 2020, arXiv e-prints,
  arXiv:2006.08662.
\newblock \doarXiv{2006.08662}

\bibitem[{{CHIME/FRB Collaboration} {et~al.}(2019{\natexlab{a}}){CHIME/FRB
  Collaboration}, {Amiri}, {Bandura}, {Bhardwaj}, {Boubel}, {Boyce}, {Boyle},
  {.~Brar}, {Burhanpurkar}, {Cassanelli}, {Chawla}, {Cliche}, {Cubranic},
  {Deng}, {Denman}, {Dobbs}, {Fandino}, {Fonseca}, {Gaensler}, {Gilbert},
  {Gill}, {Giri}, {Good}, {Halpern}, {Hanna}, {Hill}, {Hinshaw}, {H{\"o}fer},
  {Josephy}, {Kaspi}, {Landecker}, {Lang}, {Lin}, {Masui}, {Mckinven},
  {Mena-Parra}, {Merryfield}, {Michilli}, {Milutinovic}, {Moatti}, {Naidu},
  {Newburgh}, {Ng}, {Patel}, {Pen}, {Pinsonneault-Marotte}, {Pleunis},
  {Rafiei-Ravandi}, {Rahman}, {Ransom}, {Renard}, {Scholz}, {Shaw}, {Siegel},
  {Smith}, {Stairs}, {Tendulkar}, {Tretyakov}, {Vanderlinde}, \&
  {Yadav}}]{2019Natur.566..235C}
{CHIME/FRB Collaboration}, {Amiri}, M., {Bandura}, K., {et~al.}
  2019{\natexlab{a}}, \nat, 566, 235, \dodoi{10.1038/s41586-018-0864-x}

\bibitem[{{CHIME/FRB Collaboration} {et~al.}(2019{\natexlab{b}}){CHIME/FRB
  Collaboration}, {Andersen}, {Bandura}, {Bhardwaj}, {Boubel}, {Boyce},
  {Boyle}, {Brar}, {Cassanelli}, {Chawla}, {Cubranic}, {Deng}, {Dobbs},
  {Fandino}, {Fonseca}, {Gaensler}, {Gilbert}, {Giri}, {Good}, {Halpern},
  {Hill}, {Hinshaw}, {H{\"o}fer}, {Josephy}, {Kaspi}, {Kothes}, {Landecker},
  {Lang}, {Li}, {Lin}, {Masui}, {Mena-Parra}, {Merryfield}, {Mckinven},
  {Michilli}, {Milutinovic}, {Naidu}, {Newburgh}, {Ng}, {Patel}, {Pen},
  {Pinsonneault-Marotte}, {Pleunis}, {Rafiei-Ravandi}, {Rahman}, {Ransom},
  {Renard}, {Scholz}, {Siegel}, {Singh}, {Smith}, {Stairs}, {Tendulkar},
  {Tretyakov}, {Vanderlinde}, {Yadav}, \& {Zwaniga}}]{2019ApJ...885L..24C}
{CHIME/FRB Collaboration}, {Andersen}, B.~C., {Bandura}, K., {et~al.}
  2019{\natexlab{b}}, \apjl, 885, L24, \dodoi{10.3847/2041-8213/ab4a80}

\bibitem[{{CHIME/FRB Collaboration} {et~al.}(2019{\natexlab{c}}){CHIME/FRB
  Collaboration}, {Amiri}, {Bandura}, {Bhardwaj}, {Boubel}, {Boyce}, {Boyle},
  {Brar}, {Burhanpurkar}, {Chawla}, {Cliche}, {Cubranic}, {Deng}, {Denman},
  {Dobbs}, {Fand ino}, {Fonseca}, {Gaensler}, {Gilbert}, {Giri}, {Good},
  {Halpern}, {Hanna}, {Hill}, {Hinshaw}, {H{\"o}fer}, {Josephy}, {Kaspi},
  {Landecker}, {Lang}, {Masui}, {Mckinven}, {Mena-Parra}, {Merryfield},
  {Milutinovic}, {Moatti}, {Naidu}, {Newburgh}, {Ng}, {Patel}, {Pen},
  {Pinsonneault-Marotte}, {Pleunis}, {Rafiei-Ravandi}, {Ransom}, {Renard},
  {Scholz}, {Shaw}, {Siegel}, {Smith}, {Stairs}, {Tendulkar}, {Tretyakov},
  {Vand erlinde}, \& {Yadav}}]{2019Natur.566..230C}
{CHIME/FRB Collaboration}, {Amiri}, M., {Bandura}, K., {et~al.}
  2019{\natexlab{c}}, \nat, 566, 230, \dodoi{10.1038/s41586-018-0867-7}

\bibitem[{{CHIME/FRB Collaboration} {et~al.}(2020){CHIME/FRB Collaboration},
  {Andersen}, {Band ura}, {Bhardwaj}, {Bij}, {Boyce}, {Boyle}, {Brar},
  {Cassanelli}, {Chawla}, {Chen}, {Cliche}, {Cook}, {Cubranic}, {Curtin},
  {Denman}, {Dobbs}, {Dong}, {Fandino}, {Fonseca}, {Gaensler}, {Giri}, {Good},
  {Halpern}, {Hill}, {Hinshaw}, {H{\"o}fer}, {Josephy}, {Kania}, {Kaspi},
  {Landecker}, {Leung}, {Li}, {Lin}, {Masui}, {Mckinven}, {Mena-Parra},
  {Merryfield}, {Meyers}, {Michilli}, {Milutinovic}, {Mirhosseini},
  {M{\"u}nchmeyer}, {Naidu}, {Newburgh}, {Ng}, {Patel}, {Pen},
  {Pinsonneault-Marotte}, {Pleunis}, {Quine}, {Rafiei-Ravandi}, {Rahman},
  {Ransom}, {Renard}, {Sanghavi}, {Scholz}, {Shaw}, {Shin}, {Siegel}, {Singh},
  {Smegal}, {Smith}, {Stairs}, {Tan}, {Tendulkar}, {Tretyakov}, {Vanderlinde},
  {Wang}, {Wulf}, \& {Zwaniga}}]{2020arXiv200510324T}
{CHIME/FRB Collaboration}, {Andersen}, B.~C., {Band ura}, K.~M., {et~al.} 2020,
  arXiv e-prints, arXiv:2005.10324.
\newblock \doarXiv{2005.10324}

\bibitem[{{Colaiuda} \& {Kokkotas}(2012)}]{2012MNRAS.423..811C}
{Colaiuda}, A., \& {Kokkotas}, K.~D. 2012, \mnras, 423, 811,
  \dodoi{10.1111/j.1365-2966.2012.20919.x}

\bibitem[{{Collazzi} {et~al.}(2015){Collazzi}, {Kouveliotou}, {van der Horst},
  {Younes}, {Kaneko}, {G{\"o}{\u g}{\"u}{\c s}}, {Lin}, {Granot}, {Finger},
  {Chaplin}, {Huppenkothen}, {Watts}, {von Kienlin}, {Baring}, {Gruber},
  {Bhat}, {Gibby}, {Gehrels}, {McEnery}, {van der Klis}, \&
  {Wijers}}]{2015ApJS..218...11C}
{Collazzi}, A.~C., {Kouveliotou}, C., {van der Horst}, A.~J., {et~al.} 2015,
  \apjs, 218, 11, \dodoi{10.1088/0067-0049/218/1/11}

\bibitem[{{Colpi} {et~al.}(2000){Colpi}, {Geppert}, \&
  {Page}}]{2000ApJ...529L..29C}
{Colpi}, M., {Geppert}, U., \& {Page}, D. 2000, \apjl, 529, L29,
  \dodoi{10.1086/312448}

\bibitem[{{Cordes} \& {Chatterjee}(2019)}]{2019ARA&A..57..417C}
{Cordes}, J.~M., \& {Chatterjee}, S. 2019, \araa, 57, 417,
  \dodoi{10.1146/annurev-astro-091918-104501}

\bibitem[{{Cruces} {et~al.}(2020){Cruces}, {Spitler}, {Scholz}, {Lynch},
  {Seymour}, {Hessels}, {Gouiff{\`e}s}, {Hilmarsson}, {Kramer}, \&
  {Munjal}}]{2020arXiv200803461C}
{Cruces}, M., {Spitler}, L.~G., {Scholz}, P., {et~al.} 2020, arXiv e-prints,
  arXiv:2008.03461.
\newblock \doarXiv{2008.03461}

\bibitem[{{Cunningham} {et~al.}(2019){Cunningham}, {Cenko}, {Burns},
  {Goldstein}, {Lien}, {Kocevski}, {Briggs}, {Connaughton}, {Miller},
  {Racusin}, \& {Stanbro}}]{2019ApJ...879...40C}
{Cunningham}, V., {Cenko}, S.~B., {Burns}, E., {et~al.} 2019, \apj, 879, 40,
  \dodoi{10.3847/1538-4357/ab2235}

\bibitem[{{Dall'Osso} {et~al.}(2012){Dall'Osso}, {Granot}, \&
  {Piran}}]{2012MNRAS.422.2878D}
{Dall'Osso}, S., {Granot}, J., \& {Piran}, T. 2012, \mnras, 422, 2878,
  \dodoi{10.1111/j.1365-2966.2012.20612.x}

\bibitem[{{de Souza} \& {Chirenti}(2019)}]{2019PhRvD.100d3017D}
{de Souza}, G.~H., \& {Chirenti}, C. 2019, \prd, 100, 043017,
  \dodoi{10.1103/PhysRevD.100.043017}

\bibitem[{{Deibel} {et~al.}(2014){Deibel}, {Steiner}, \&
  {Brown}}]{2014PhRvC..90b5802D}
{Deibel}, A.~T., {Steiner}, A.~W., \& {Brown}, E.~F. 2014, \prc, 90, 025802,
  \dodoi{10.1103/PhysRevC.90.025802}

\bibitem[{{Douchin} \& {Haensel}(2001)}]{DouchinHaensel2001}
{Douchin}, F., \& {Haensel}, P. 2001, \aap, 380, 151,
  \dodoi{10.1051/0004-6361:20011402}

\bibitem[{{Duncan}(1998)}]{1998ApJ...498L..45D}
{Duncan}, R.~C. 1998, \apjl, 498, L45, \dodoi{10.1086/311303}

\bibitem[{{Fonseca} {et~al.}(2020){Fonseca}, {Andersen}, {Bhardwaj}, {Chawla},
  {Good}, {Josephy}, {Kaspi}, {Masui}, {Mckinven}, {Michilli}, {Pleunis},
  {Shin}, {Tendulkar}, {Bandura}, {Boyle}, {Brar}, {Cassanelli}, {Cubranic},
  {Dobbs}, {Dong}, {Gaensler}, {Hinshaw}, {Land ecker}, {Leung}, {Li}, {Lin},
  {Mena-Parra}, {Merryfield}, {Naidu}, {Ng}, {Patel}, {Pen}, {Rafiei-Ravandi},
  {Rahman}, {Ransom}, {Scholz}, {Smith}, {Stairs}, {Vanderlinde}, {Yadav}, \&
  {Zwaniga}}]{2020ApJ...891L...6F}
{Fonseca}, E., {Andersen}, B.~C., {Bhardwaj}, M., {et~al.} 2020, \apjl, 891,
  L6, \dodoi{10.3847/2041-8213/ab7208}

\bibitem[{{Gabler} {et~al.}(2012){Gabler}, {Cerd{\'a}-Dur{\'a}n},
  {Stergioulas}, {Font}, \& {M{\"u}ller}}]{2012MNRAS.421.2054G}
{Gabler}, M., {Cerd{\'a}-Dur{\'a}n}, P., {Stergioulas}, N., {Font}, J.~A., \&
  {M{\"u}ller}, E. 2012, \mnras, 421, 2054,
  \dodoi{10.1111/j.1365-2966.2012.20454.x}

\bibitem[{{Gourdji} {et~al.}(2019){Gourdji}, {Michilli}, {Spitler}, {Hessels},
  {Seymour}, {Cordes}, \& {Chatterjee}}]{2019ApJ...877L..19G}
{Gourdji}, K., {Michilli}, D., {Spitler}, L.~G., {et~al.} 2019, \apjl, 877,
  L19, \dodoi{10.3847/2041-8213/ab1f8a}

\bibitem[{{Hardy} {et~al.}(2017){Hardy}, {Dhillon}, {Spitler}, {Littlefair},
  {Ashley}, {De Cia}, {Green}, {Jaroenjittichai}, {Keane}, {Kerry}, {Kramer},
  {Malesani}, {Marsh}, {Parsons}, {Possenti}, {Rattanasoon}, \&
  {Sahman}}]{2017MNRAS.472.2800H}
{Hardy}, L.~K., {Dhillon}, V.~S., {Spitler}, L.~G., {et~al.} 2017, \mnras, 472,
  2800, \dodoi{10.1093/mnras/stx2153}

\bibitem[{{Hessels} {et~al.}(2019){Hessels}, {Spitler}, {Seymour}, {Cordes},
  {Michilli}, {Lynch}, {Gourdji}, {Archibald}, {Bassa}, {Bower}, {Chatterjee},
  {Connor}, {Crawford}, {Deneva}, {Gajjar}, {Kaspi}, {Keimpema}, {Law},
  {Marcote}, {McLaughlin}, {Paragi}, {Petroff}, {Ransom}, {Scholz}, {Stappers},
  \& {Tendulkar}}]{2019ApJ...876L..23H}
{Hessels}, J.~W.~T., {Spitler}, L.~G., {Seymour}, A.~D., {et~al.} 2019, The
  Astrophysical Journal, 876, L23, \dodoi{10.3847/2041-8213/ab13ae}

\bibitem[{{Huppenkothen} {et~al.}(2014{\natexlab{a}}){Huppenkothen}, {Heil},
  {Watts}, \& {G{\"o}{\u{g}}{\"u}{\textcommabelow s}}}]{2014ApJ...795..114H}
{Huppenkothen}, D., {Heil}, L.~M., {Watts}, A.~L., \&
  {G{\"o}{\u{g}}{\"u}{\textcommabelow s}}, E. 2014{\natexlab{a}}, \apj, 795,
  114, \dodoi{10.1088/0004-637X/795/2/114}

\bibitem[{{Huppenkothen} {et~al.}(2014{\natexlab{b}}){Huppenkothen}, {Watts},
  \& {Levin}}]{2014ApJ...793..129H}
{Huppenkothen}, D., {Watts}, A.~L., \& {Levin}, Y. 2014{\natexlab{b}}, \apj,
  793, 129, \dodoi{10.1088/0004-637X/793/2/129}

\bibitem[{{Huppenkothen} {et~al.}(2014{\natexlab{c}}){Huppenkothen},
  {D'Angelo}, {Watts}, {Heil}, {van der Klis}, {van der Horst}, {Kouveliotou},
  {Baring}, {G{\"o}{\u g}{\"u}{\c s}}, {Granot}, {Kaneko}, {Lin}, {von
  Kienlin}, \& {Younes}}]{2014ApJ...787..128H}
{Huppenkothen}, D., {D'Angelo}, C., {Watts}, A.~L., {et~al.}
  2014{\natexlab{c}}, \apj, 787, 128, \dodoi{10.1088/0004-637X/787/2/128}

\bibitem[{{Huppenkothen} {et~al.}(2015){Huppenkothen}, {Brewer}, {Hogg},
  {Murray}, {Frean}, {Elenbaas}, {Watts}, {Levin}, {van der Horst}, \&
  {Kouveliotou}}]{2015ApJ...810...66H}
{Huppenkothen}, D., {Brewer}, B.~J., {Hogg}, D.~W., {et~al.} 2015, \apj, 810,
  66, \dodoi{10.1088/0004-637X/810/1/66}

\bibitem[{{Israel} {et~al.}(2005){Israel}, {Belloni}, {Stella}, {Rephaeli},
  {Gruber}, {Casella}, {Dall'Osso}, {Rea}, {Persic}, \&
  {Rothschild}}]{2005ApJ...628L..53I}
{Israel}, G.~L., {Belloni}, T., {Stella}, L., {et~al.} 2005, \apjl, 628, L53,
  \dodoi{10.1086/432615}

\bibitem[{{Kiziltan} {et~al.}(2013){Kiziltan}, {Kottas}, {De Yoreo}, \&
  {Thorsett}}]{2013ApJ...778...66K}
{Kiziltan}, B., {Kottas}, A., {De Yoreo}, M., \& {Thorsett}, S.~E. 2013, \apj,
  778, 66, \dodoi{10.1088/0004-637X/778/1/66}

\bibitem[{{Kumar} {et~al.}(2017){Kumar}, {Lu}, \&
  {Bhattacharya}}]{2017MNRAS.468.2726K}
{Kumar}, P., {Lu}, W., \& {Bhattacharya}, M. 2017, \mnras, 468, 2726,
  \dodoi{10.1093/mnras/stx665}

\bibitem[{{Laha}(2020)}]{2020PhRvD.102b3016L}
{Laha}, R. 2020, \prd, 102, 023016, \dodoi{10.1103/PhysRevD.102.023016}

\bibitem[{{Levin}(2006)}]{2006MNRAS.368L..35L}
{Levin}, Y. 2006, \mnras, 368, L35, \dodoi{10.1111/j.1745-3933.2006.00155.x}

\bibitem[{{Levin} \& {Lyutikov}(2012)}]{2012MNRAS.427.1574L}
{Levin}, Y., \& {Lyutikov}, M. 2012, \mnras, 427, 1574,
  \dodoi{10.1111/j.1365-2966.2012.22016.x}

\bibitem[{{Li} {et~al.}(2020){Li}, {Lin}, {Xiong}, {Ge}, {Li}, {Li}, {Lu},
  {Zhang}, {Tuo}, {Nang}, {Zhang}, {Xiao}, {Chen}, {Song}, {Xu}, {Liu}, {Jia},
  {Cao}, {Zhang}, {Qu}, {Liao}, {Zhao}, {Tan}, {Nie}, {Zhao}, {Zheng}, {Zheng},
  {Luo}, {Cai}, {Li}, {Xue}, {Bu}, {Chang}, {Chen}, {Chen}, {Chen}, {Chen},
  {Chen}, {Cui}, {Cui}, {Deng}, {Dong}, {Du}, {Fu}, {Gao}, {Gao}, {Gao}, {Gu},
  {Guan}, {Guo}, {Han}, {Huang}, {Huo}, {Jiang}, {Jiang}, {Jin}, {Jin}, {Kong},
  {Li}, {Li}, {Li}, {Li}, {Li}, {Li}, {Li}, {Liang}, {Liu}, {Liu}, {Liu},
  {Liu}, {Liu}, {Lu}, {Lu}, {Luo}, {Ma}, {Meng}, {Ou}, {Sai}, {Shang}, {Song},
  {Sun}, {Tao}, {Wang}, {Wang}, {Wang}, {Wang}, {Wang}, {Wen}, {Wu}, {Wu},
  {Wu}, {Xiao}, {Yang}, {Yang}, {Yang}, {Yang}, {Yi}, {Yin}, {You}, {Zhang},
  {Zhang}, {Zhang}, {Zhang}, {Zhang}, {Zhang}, {Zhang}, {Zhang}, {Zhang},
  {Zhang}, {Zhang}, {Zhang}, {Zhang}, {Zhang}, {Zhang}, {Zhang}, {Zhou},
  {Zhou}, {Zhu}, {Zhu}, \& {Zhuang}}]{2020arXiv200511071L}
{Li}, C.~K., {Lin}, L., {Xiong}, S.~L., {et~al.} 2020, arXiv e-prints,
  arXiv:2005.11071.
\newblock \doarXiv{2005.11071}

\bibitem[{{Li} {et~al.}(2018){Li}, {Gao}, {Ding}, {Wang}, \&
  {Zhang}}]{2018NatCo...9.3833L}
{Li}, Z.-X., {Gao}, H., {Ding}, X.-H., {Wang}, G.-J., \& {Zhang}, B. 2018,
  Nature Communications, 9, 3833, \dodoi{10.1038/s41467-018-06303-0}

\bibitem[{{Lin} {et~al.}(2012){Lin}, {G{\"o}{\v{g}}{\"u}{\textcommabelow s}},
  {Baring}, {Granot}, {Kouveliotou}, {Kaneko}, {van der Horst}, {Gruber}, {von
  Kienlin}, {Younes}, {Watts}, \& {Gehrels}}]{2012ApJ...756...54L}
{Lin}, L., {G{\"o}{\v{g}}{\"u}{\textcommabelow s}}, E., {Baring}, M.~G.,
  {et~al.} 2012, \apj, 756, 54, \dodoi{10.1088/0004-637X/756/1/54}

\bibitem[{{Link} \& {van Eysden}(2016)}]{2016ApJ...823L...1L}
{Link}, B., \& {van Eysden}, C.~A. 2016, \apjl, 823, L1,
  \dodoi{10.3847/2041-8205/823/1/L1}

\bibitem[{{Lyutikov}(2017)}]{2017ApJ...838L..13L}
{Lyutikov}, M. 2017, \apjl, 838, L13, \dodoi{10.3847/2041-8213/aa62fa}

\bibitem[{{Lyutikov} \& {Popov}(2020)}]{2020arXiv200505093L}
{Lyutikov}, M., \& {Popov}, S. 2020, arXiv e-prints, arXiv:2005.05093.
\newblock \doarXiv{2005.05093}

\bibitem[{{Macquart} {et~al.}(2020){Macquart}, {Prochaska}, {McQuinn},
  {Bannister}, {Bhandari}, {Day}, {Deller}, {Ekers}, {James}, {Marnoch},
  {Os{\l}owski}, {Phillips}, {Ryder}, {Scott}, {Shannon}, \&
  {Tejos}}]{2020Natur.581..391M}
{Macquart}, J.~P., {Prochaska}, J.~X., {McQuinn}, M., {et~al.} 2020, \nat, 581,
  391, \dodoi{10.1038/s41586-020-2300-2}

\bibitem[{{Margalit} {et~al.}(2020){Margalit}, {Beniamini}, {Sridhar}, \&
  {Metzger}}]{2020arXiv200505283M}
{Margalit}, B., {Beniamini}, P., {Sridhar}, N., \& {Metzger}, B.~D. 2020, arXiv
  e-prints, arXiv:2005.05283.
\newblock \doarXiv{2005.05283}

\bibitem[{{Mereghetti}(2008)}]{2008A&ARv..15..225M}
{Mereghetti}, S. 2008, \aapr, 15, 225, \dodoi{10.1007/s00159-008-0011-z}

\bibitem[{{Mereghetti} {et~al.}(2020){Mereghetti}, {Savchenko}, {Ferrigno},
  {G{\"o}tz}, {Rigoselli}, {Tiengo}, {Bazzano}, {Bozzo}, {Coleiro},
  {Courvoisier}, {Doyle}, {Goldwurm}, {Hanlon}, {Jourdain}, {von Kienlin},
  {Lutovinov}, {Martin-Carrillo}, {Molkov}, {Natalucci}, {Onori}, {Panessa},
  {Rodi}, {Rodriguez}, {S{\'a}nchez-Fern{\'a}ndez}, {Sunyaev}, \&
  {Ubertini}}]{2020arXiv200506335M}
{Mereghetti}, S., {Savchenko}, V., {Ferrigno}, C., {et~al.} 2020, arXiv
  e-prints, arXiv:2005.06335.
\newblock \doarXiv{2005.06335}

\bibitem[{{Messios} {et~al.}(2001){Messios}, {Papadopoulos}, \&
  {Stergioulas}}]{2001MNRAS.328.1161M}
{Messios}, N., {Papadopoulos}, D.~B., \& {Stergioulas}, N. 2001, \mnras, 328,
  1161, \dodoi{10.1046/j.1365-8711.2001.04645.x}

\bibitem[{{Miller} {et~al.}(2020){Miller}, {Chirenti}, \&
  {Lamb}}]{2020ApJ...888...12M}
{Miller}, M.~C., {Chirenti}, C., \& {Lamb}, F.~K. 2020, \apj, 888, 12,
  \dodoi{10.3847/1538-4357/ab4ef9}

\bibitem[{{Miller} {et~al.}(2019{\natexlab{a}}){Miller}, {Chirenti}, \&
  {Strohmayer}}]{2019ApJ...871...95M}
{Miller}, M.~C., {Chirenti}, C., \& {Strohmayer}, T.~E. 2019{\natexlab{a}},
  \apj, 871, 95, \dodoi{10.3847/1538-4357/aaf5ce}

\bibitem[{{Miller} {et~al.}(2019{\natexlab{b}}){Miller}, {Lamb}, {Dittmann},
  {Bogdanov}, {Arzoumanian}, {Gendreau}, {Guillot}, {Harding}, {Ho},
  {Lattimer}, {Ludlam}, {Mahmoodifar}, {Morsink}, {Ray}, {Strohmayer}, {Wood},
  {Enoto}, {Foster}, {Okajima}, {Prigozhin}, \& {Soong}}]{2019ApJ...887L..24M}
{Miller}, M.~C., {Lamb}, F.~K., {Dittmann}, A.~J., {et~al.} 2019{\natexlab{b}},
  \apjl, 887, L24, \dodoi{10.3847/2041-8213/ab50c5}

\bibitem[{{Mu{\~n}oz} {et~al.}(2020){Mu{\~n}oz}, {Ravi}, \&
  {Loeb}}]{2020ApJ...890..162M}
{Mu{\~n}oz}, J.~B., {Ravi}, V., \& {Loeb}, A. 2020, \apj, 890, 162,
  \dodoi{10.3847/1538-4357/ab6d62}

\bibitem[{{{\"O}zel} {et~al.}(2012){{\"O}zel}, {Psaltis}, {Narayan}, \& {Santos
  Villarreal}}]{2012ApJ...757...55O}
{{\"O}zel}, F., {Psaltis}, D., {Narayan}, R., \& {Santos Villarreal}, A. 2012,
  \apj, 757, 55, \dodoi{10.1088/0004-637X/757/1/55}

\bibitem[{{Perna} \& {Pons}(2011)}]{2011ApJ...727L..51P}
{Perna}, R., \& {Pons}, J.~A. 2011, \apj, 727, L51,
  \dodoi{10.1088/2041-8205/727/2/L51}

\bibitem[{Philippov {et~al.}(2020)Philippov, Timokhin, \&
  Spitkovsky}]{PhysRevLett.124.245101}
Philippov, A., Timokhin, A., \& Spitkovsky, A. 2020, Phys. Rev. Lett., 124,
  245101, \dodoi{10.1103/PhysRevLett.124.245101}

\bibitem[{{Riley} {et~al.}(2019){Riley}, {Watts}, {Bogdanov}, {Ray}, {Ludlam},
  {Guillot}, {Arzoumanian}, {Baker}, {Bilous}, {Chakrabarty}, {Gendreau},
  {Harding}, {Ho}, {Lattimer}, {Morsink}, \&
  {Strohmayer}}]{2019ApJ...887L..21R}
{Riley}, T.~E., {Watts}, A.~L., {Bogdanov}, S., {et~al.} 2019, \apjl, 887, L21,
  \dodoi{10.3847/2041-8213/ab481c}

\bibitem[{{Sammons} {et~al.}(2020){Sammons}, {Macquart}, {Ekers}, {Shannon},
  {Cho}, {Prochaska}, {Deller}, \& {Day}}]{2020arXiv200212533S}
{Sammons}, M.~W., {Macquart}, J.-P., {Ekers}, R.~D., {et~al.} 2020, arXiv
  e-prints, arXiv:2002.12533.
\newblock \doarXiv{2002.12533}

\bibitem[{{Samuelsson} \& {Andersson}(2007)}]{2007MNRAS.374..256S}
{Samuelsson}, L., \& {Andersson}, N. 2007, \mnras, 374, 256,
  \dodoi{10.1111/j.1365-2966.2006.11147.x}

\bibitem[{{Simard} \& {Ravi}(2020)}]{2020arXiv200613184S}
{Simard}, D., \& {Ravi}, V. 2020, arXiv e-prints, arXiv:2006.13184.
\newblock \doarXiv{2006.13184}

\bibitem[{{Sotani} {et~al.}(2016){Sotani}, {Iida}, \&
  {Oyamatsu}}]{2016NewA...43...80S}
{Sotani}, H., {Iida}, K., \& {Oyamatsu}, K. 2016, \na, 43, 80,
  \dodoi{10.1016/j.newast.2015.08.003}

\bibitem[{{Sotani} {et~al.}(2007){Sotani}, {Kokkotas}, \&
  {Stergioulas}}]{2007MNRAS.375..261S}
{Sotani}, H., {Kokkotas}, K.~D., \& {Stergioulas}, N. 2007, \mnras, 375, 261,
  \dodoi{10.1111/j.1365-2966.2006.11304.x}

\bibitem[{{Steiner}(2012)}]{2012PhRvC..85e5804S}
{Steiner}, A.~W. 2012, \prc, 85, 055804, \dodoi{10.1103/PhysRevC.85.055804}

\bibitem[{{Strohmayer} \& {Watts}(2005)}]{2005ApJ...632L.111S}
{Strohmayer}, T.~E., \& {Watts}, A.~L. 2005, \apjl, 632, L111,
  \dodoi{10.1086/497911}

\bibitem[{{Strohmayer} \& {Watts}(2006)}]{2006ApJ...653..593S}
---. 2006, \apj, 653, 593, \dodoi{10.1086/508703}

\bibitem[{{Suvorov} \& {Kokkotas}(2019)}]{2019MNRAS.488.5887S}
{Suvorov}, A.~G., \& {Kokkotas}, K.~D. 2019, \mnras, 488, 5887,
  \dodoi{10.1093/mnras/stz2052}

\bibitem[{{Tendulkar} {et~al.}(2017){Tendulkar}, {Bassa}, {Cordes}, {Bower},
  {Law}, {Chatterjee}, {Adams}, {Bogdanov}, {Burke-Spolaor}, {Butler},
  {Demorest}, {Hessels}, {Kaspi}, {Lazio}, {Maddox}, {Marcote}, {McLaughlin},
  {Paragi}, {Ransom}, {Scholz}, {Seymour}, {Spitler}, {van Langevelde}, \&
  {Wharton}}]{2017ApJ...834L...7T}
{Tendulkar}, S.~P., {Bassa}, C.~G., {Cordes}, J.~M., {et~al.} 2017, \apjl, 834,
  L7, \dodoi{10.3847/2041-8213/834/2/L7}

\bibitem[{{Thompson} {et~al.}(2017){Thompson}, {Yang}, \&
  {Ortiz}}]{2017ApJ...841...54T}
{Thompson}, C., {Yang}, H., \& {Ortiz}, N. 2017, \apj, 841, 54,
  \dodoi{10.3847/1538-4357/aa6c30}

\bibitem[{{Timokhin}(2010)}]{2010MNRAS.408.2092T}
{Timokhin}, A.~N. 2010, \mnras, 408, 2092,
  \dodoi{10.1111/j.1365-2966.2010.17286.x}

\bibitem[{{Timokhin} \& {Arons}(2013)}]{2013MNRAS.429...20T}
{Timokhin}, A.~N., \& {Arons}, J. 2013, \mnras, 429, 20,
  \dodoi{10.1093/mnras/sts298}

\bibitem[{{Timokhin} {et~al.}(2008){Timokhin}, {Eichler}, \&
  {Lyubarsky}}]{2008ApJ...680.1398T}
{Timokhin}, A.~N., {Eichler}, D., \& {Lyubarsky}, Y. 2008, \apj, 680, 1398,
  \dodoi{10.1086/587925}

\bibitem[{{Turolla} {et~al.}(2015){Turolla}, {Zane}, \&
  {Watts}}]{2015RPPh...78k6901T}
{Turolla}, R., {Zane}, S., \& {Watts}, A.~L. 2015, Reports on Progress in
  Physics, 78, 116901, \dodoi{10.1088/0034-4885/78/11/116901}

\bibitem[{{Wadiasingh} {et~al.}(2020){Wadiasingh}, {Beniamini}, {Timokhin},
  {Baring}, {van der Horst}, {Harding}, \& {Kazanas}}]{Wadiasingh2020}
{Wadiasingh}, Z., {Beniamini}, P., {Timokhin}, A., {et~al.} 2020, \apj, 891,
  82, \dodoi{10.3847/1538-4357/ab6d69}

\bibitem[{{Wadiasingh} \& {Timokhin}(2019)}]{2019ApJ...879....4W}
{Wadiasingh}, Z., \& {Timokhin}, A. 2019, \apj, 879, 4,
  \dodoi{10.3847/1538-4357/ab2240}

\bibitem[{{Wang} {et~al.}(2018){Wang}, {Luo}, {Yue}, {Chen}, {Lee}, \&
  {Xu}}]{2018ApJ...852..140W}
{Wang}, W., {Luo}, R., {Yue}, H., {et~al.} 2018, \apj, 852, 140,
  \dodoi{10.3847/1538-4357/aaa025}

\bibitem[{{Watts} \& {Strohmayer}(2006)}]{2006ApJ...637L.117W}
{Watts}, A.~L., \& {Strohmayer}, T.~E. 2006, \apjl, 637, L117,
  \dodoi{10.1086/500735}

\bibitem[{{Watts} \& {Strohmayer}(2007)}]{2007AdSpR..40.1446W}
---. 2007, Advances in Space Research, 40, 1446,
  \dodoi{10.1016/j.asr.2006.12.021}

\bibitem[{{Younes} {et~al.}(2020){Younes}, {Baring}, {Kouveliotou},
  {Arzoumanian}, {Enoto}, {Doty}, {Gendreau},
  {G{\"o}{\u{g}}{\"u}{\textcommabelow s}}, {Guillot}, {G{\"u}ver}, {Harding},
  {Ho}, {van der Horst}, {Jaisawal}, {Kaneko}, {LaMarr}, {Lin}, {Majid},
  {Okajima}, {Pope}, {Ray}, {Roberts}, {Saylor}, {Steiner}, \&
  {Wadiasingh}}]{2020arXiv200611358Y}
{Younes}, G., {Baring}, M.~G., {Kouveliotou}, C., {et~al.} 2020, arXiv
  e-prints, arXiv:2006.11358.
\newblock \doarXiv{2006.11358}

\bibitem[{{Zhang} {et~al.}(2018){Zhang}, {Gajjar}, {Foster}, {Siemion},
  {Cordes}, {Law}, \& {Wang}}]{2018ApJ...866..149Z}
{Zhang}, Y.~G., {Gajjar}, V., {Foster}, G., {et~al.} 2018, \apj, 866, 149,
  \dodoi{10.3847/1538-4357/aadf31}

\end{thebibliography}

\end{document}